\documentclass[letterpaper,compsoc,twoside]{IEEEtran}
\usepackage{fixltx2e} % LaTeX patches, \textsubscript
\usepackage{cmap} % fix search and cut-and-paste in Acrobat
\usepackage{ifthen}
\usepackage[T1]{fontenc}
\usepackage[utf8]{inputenc}
\usepackage{amsmath}
\usepackage{color}

\usepackage[font={small,it},labelfont=bf]{caption}
\usepackage{float}

\usepackage{scipy}
\makeatletter
\def\PY@reset{\let\PY@it=\relax \let\PY@bf=\relax%
    \let\PY@ul=\relax \let\PY@tc=\relax%
    \let\PY@bc=\relax \let\PY@ff=\relax}
\def\PY@tok#1{\csname PY@tok@#1\endcsname}
\def\PY@toks#1+{\ifx\relax#1\empty\else%
    \PY@tok{#1}\expandafter\PY@toks\fi}
\def\PY@do#1{\PY@bc{\PY@tc{\PY@ul{%
    \PY@it{\PY@bf{\PY@ff{#1}}}}}}}
\def\PY#1#2{\PY@reset\PY@toks#1+\relax+\PY@do{#2}}

\expandafter\def\csname PY@tok@gd\endcsname{\def\PY@tc##1{\textcolor[rgb]{0.63,0.00,0.00}{##1}}}
\expandafter\def\csname PY@tok@gu\endcsname{\let\PY@bf=\textbf\def\PY@tc##1{\textcolor[rgb]{0.50,0.00,0.50}{##1}}}
\expandafter\def\csname PY@tok@gt\endcsname{\def\PY@tc##1{\textcolor[rgb]{0.00,0.27,0.87}{##1}}}
\expandafter\def\csname PY@tok@gs\endcsname{\let\PY@bf=\textbf}
\expandafter\def\csname PY@tok@gr\endcsname{\def\PY@tc##1{\textcolor[rgb]{1.00,0.00,0.00}{##1}}}
\expandafter\def\csname PY@tok@cm\endcsname{\let\PY@it=\textit\def\PY@tc##1{\textcolor[rgb]{0.25,0.50,0.56}{##1}}}
\expandafter\def\csname PY@tok@vg\endcsname{\def\PY@tc##1{\textcolor[rgb]{0.73,0.38,0.84}{##1}}}
\expandafter\def\csname PY@tok@vi\endcsname{\def\PY@tc##1{\textcolor[rgb]{0.73,0.38,0.84}{##1}}}
\expandafter\def\csname PY@tok@vm\endcsname{\def\PY@tc##1{\textcolor[rgb]{0.73,0.38,0.84}{##1}}}
\expandafter\def\csname PY@tok@mh\endcsname{\def\PY@tc##1{\textcolor[rgb]{0.13,0.50,0.31}{##1}}}
\expandafter\def\csname PY@tok@cs\endcsname{\def\PY@tc##1{\textcolor[rgb]{0.25,0.50,0.56}{##1}}\def\PY@bc##1{\setlength{\fboxsep}{0pt}\colorbox[rgb]{1.00,0.94,0.94}{\strut ##1}}}
\expandafter\def\csname PY@tok@ge\endcsname{\let\PY@it=\textit}
\expandafter\def\csname PY@tok@vc\endcsname{\def\PY@tc##1{\textcolor[rgb]{0.73,0.38,0.84}{##1}}}
\expandafter\def\csname PY@tok@il\endcsname{\def\PY@tc##1{\textcolor[rgb]{0.13,0.50,0.31}{##1}}}
\expandafter\def\csname PY@tok@go\endcsname{\def\PY@tc##1{\textcolor[rgb]{0.20,0.20,0.20}{##1}}}
\expandafter\def\csname PY@tok@cp\endcsname{\def\PY@tc##1{\textcolor[rgb]{0.00,0.44,0.13}{##1}}}
\expandafter\def\csname PY@tok@gi\endcsname{\def\PY@tc##1{\textcolor[rgb]{0.00,0.63,0.00}{##1}}}
\expandafter\def\csname PY@tok@gh\endcsname{\let\PY@bf=\textbf\def\PY@tc##1{\textcolor[rgb]{0.00,0.00,0.50}{##1}}}
\expandafter\def\csname PY@tok@ni\endcsname{\let\PY@bf=\textbf\def\PY@tc##1{\textcolor[rgb]{0.84,0.33,0.22}{##1}}}
\expandafter\def\csname PY@tok@nl\endcsname{\let\PY@bf=\textbf\def\PY@tc##1{\textcolor[rgb]{0.00,0.13,0.44}{##1}}}
\expandafter\def\csname PY@tok@nn\endcsname{\let\PY@bf=\textbf\def\PY@tc##1{\textcolor[rgb]{0.05,0.52,0.71}{##1}}}
\expandafter\def\csname PY@tok@no\endcsname{\def\PY@tc##1{\textcolor[rgb]{0.38,0.68,0.84}{##1}}}
\expandafter\def\csname PY@tok@na\endcsname{\def\PY@tc##1{\textcolor[rgb]{0.25,0.44,0.63}{##1}}}
\expandafter\def\csname PY@tok@nb\endcsname{\def\PY@tc##1{\textcolor[rgb]{0.00,0.44,0.13}{##1}}}
\expandafter\def\csname PY@tok@nc\endcsname{\let\PY@bf=\textbf\def\PY@tc##1{\textcolor[rgb]{0.05,0.52,0.71}{##1}}}
\expandafter\def\csname PY@tok@nd\endcsname{\let\PY@bf=\textbf\def\PY@tc##1{\textcolor[rgb]{0.33,0.33,0.33}{##1}}}
\expandafter\def\csname PY@tok@ne\endcsname{\def\PY@tc##1{\textcolor[rgb]{0.00,0.44,0.13}{##1}}}
\expandafter\def\csname PY@tok@nf\endcsname{\def\PY@tc##1{\textcolor[rgb]{0.02,0.16,0.49}{##1}}}
\expandafter\def\csname PY@tok@si\endcsname{\let\PY@it=\textit\def\PY@tc##1{\textcolor[rgb]{0.44,0.63,0.82}{##1}}}
\expandafter\def\csname PY@tok@s2\endcsname{\def\PY@tc##1{\textcolor[rgb]{0.25,0.44,0.63}{##1}}}
\expandafter\def\csname PY@tok@nt\endcsname{\let\PY@bf=\textbf\def\PY@tc##1{\textcolor[rgb]{0.02,0.16,0.45}{##1}}}
\expandafter\def\csname PY@tok@nv\endcsname{\def\PY@tc##1{\textcolor[rgb]{0.73,0.38,0.84}{##1}}}
\expandafter\def\csname PY@tok@s1\endcsname{\def\PY@tc##1{\textcolor[rgb]{0.25,0.44,0.63}{##1}}}
\expandafter\def\csname PY@tok@dl\endcsname{\def\PY@tc##1{\textcolor[rgb]{0.25,0.44,0.63}{##1}}}
\expandafter\def\csname PY@tok@ch\endcsname{\let\PY@it=\textit\def\PY@tc##1{\textcolor[rgb]{0.25,0.50,0.56}{##1}}}
\expandafter\def\csname PY@tok@m\endcsname{\def\PY@tc##1{\textcolor[rgb]{0.13,0.50,0.31}{##1}}}
\expandafter\def\csname PY@tok@gp\endcsname{\let\PY@bf=\textbf\def\PY@tc##1{\textcolor[rgb]{0.78,0.36,0.04}{##1}}}
\expandafter\def\csname PY@tok@sh\endcsname{\def\PY@tc##1{\textcolor[rgb]{0.25,0.44,0.63}{##1}}}
\expandafter\def\csname PY@tok@ow\endcsname{\let\PY@bf=\textbf\def\PY@tc##1{\textcolor[rgb]{0.00,0.44,0.13}{##1}}}
\expandafter\def\csname PY@tok@sx\endcsname{\def\PY@tc##1{\textcolor[rgb]{0.78,0.36,0.04}{##1}}}
\expandafter\def\csname PY@tok@bp\endcsname{\def\PY@tc##1{\textcolor[rgb]{0.00,0.44,0.13}{##1}}}
\expandafter\def\csname PY@tok@c1\endcsname{\let\PY@it=\textit\def\PY@tc##1{\textcolor[rgb]{0.25,0.50,0.56}{##1}}}
\expandafter\def\csname PY@tok@fm\endcsname{\def\PY@tc##1{\textcolor[rgb]{0.02,0.16,0.49}{##1}}}
\expandafter\def\csname PY@tok@o\endcsname{\def\PY@tc##1{\textcolor[rgb]{0.40,0.40,0.40}{##1}}}
\expandafter\def\csname PY@tok@kc\endcsname{\let\PY@bf=\textbf\def\PY@tc##1{\textcolor[rgb]{0.00,0.44,0.13}{##1}}}
\expandafter\def\csname PY@tok@c\endcsname{\let\PY@it=\textit\def\PY@tc##1{\textcolor[rgb]{0.25,0.50,0.56}{##1}}}
\expandafter\def\csname PY@tok@mf\endcsname{\def\PY@tc##1{\textcolor[rgb]{0.13,0.50,0.31}{##1}}}
\expandafter\def\csname PY@tok@err\endcsname{\def\PY@bc##1{\setlength{\fboxsep}{0pt}\fcolorbox[rgb]{1.00,0.00,0.00}{1,1,1}{\strut ##1}}}
\expandafter\def\csname PY@tok@mb\endcsname{\def\PY@tc##1{\textcolor[rgb]{0.13,0.50,0.31}{##1}}}
\expandafter\def\csname PY@tok@ss\endcsname{\def\PY@tc##1{\textcolor[rgb]{0.32,0.47,0.09}{##1}}}
\expandafter\def\csname PY@tok@sr\endcsname{\def\PY@tc##1{\textcolor[rgb]{0.14,0.33,0.53}{##1}}}
\expandafter\def\csname PY@tok@mo\endcsname{\def\PY@tc##1{\textcolor[rgb]{0.13,0.50,0.31}{##1}}}
\expandafter\def\csname PY@tok@kd\endcsname{\let\PY@bf=\textbf\def\PY@tc##1{\textcolor[rgb]{0.00,0.44,0.13}{##1}}}
\expandafter\def\csname PY@tok@mi\endcsname{\def\PY@tc##1{\textcolor[rgb]{0.13,0.50,0.31}{##1}}}
\expandafter\def\csname PY@tok@kn\endcsname{\let\PY@bf=\textbf\def\PY@tc##1{\textcolor[rgb]{0.00,0.44,0.13}{##1}}}
\expandafter\def\csname PY@tok@cpf\endcsname{\let\PY@it=\textit\def\PY@tc##1{\textcolor[rgb]{0.25,0.50,0.56}{##1}}}
\expandafter\def\csname PY@tok@kr\endcsname{\let\PY@bf=\textbf\def\PY@tc##1{\textcolor[rgb]{0.00,0.44,0.13}{##1}}}
\expandafter\def\csname PY@tok@s\endcsname{\def\PY@tc##1{\textcolor[rgb]{0.25,0.44,0.63}{##1}}}
\expandafter\def\csname PY@tok@kp\endcsname{\def\PY@tc##1{\textcolor[rgb]{0.00,0.44,0.13}{##1}}}
\expandafter\def\csname PY@tok@w\endcsname{\def\PY@tc##1{\textcolor[rgb]{0.73,0.73,0.73}{##1}}}
\expandafter\def\csname PY@tok@kt\endcsname{\def\PY@tc##1{\textcolor[rgb]{0.56,0.13,0.00}{##1}}}
\expandafter\def\csname PY@tok@sc\endcsname{\def\PY@tc##1{\textcolor[rgb]{0.25,0.44,0.63}{##1}}}
\expandafter\def\csname PY@tok@sb\endcsname{\def\PY@tc##1{\textcolor[rgb]{0.25,0.44,0.63}{##1}}}
\expandafter\def\csname PY@tok@sa\endcsname{\def\PY@tc##1{\textcolor[rgb]{0.25,0.44,0.63}{##1}}}
\expandafter\def\csname PY@tok@k\endcsname{\let\PY@bf=\textbf\def\PY@tc##1{\textcolor[rgb]{0.00,0.44,0.13}{##1}}}
\expandafter\def\csname PY@tok@se\endcsname{\let\PY@bf=\textbf\def\PY@tc##1{\textcolor[rgb]{0.25,0.44,0.63}{##1}}}
\expandafter\def\csname PY@tok@sd\endcsname{\let\PY@it=\textit\def\PY@tc##1{\textcolor[rgb]{0.25,0.44,0.63}{##1}}}

\def\PYZus{\char`\_}
\def\PYZob{\char`\{}
\def\PYZcb{\char`\}}

\def\PYZgt{\char`\>}
\def\PYZsh{\char`\#}
\def\PYZpc{\char`\%}

\def\PYZhy{\char`\-}
\def\PYZsq{\char`\'}

\makeatother

\providecommand*{\DUfootnotemark}[3]{%
  \raisebox{1em}{\hypertarget{#1}{}}%
  \hyperlink{#2}{\textsuperscript{#3}}%
}
\providecommand{\DUfootnotetext}[4]{%
  \begingroup%
  \renewcommand{\thefootnote}{%
    \protect\raisebox{1em}{\protect\hypertarget{#1}{}}%
    \protect\hyperlink{#2}{#3}}%
  \footnotetext{#4}%
  \endgroup%
}

\providecommand*{\DUrole}[2]{%
  \ifcsname DUrole#1\endcsname%
    \csname DUrole#1\endcsname{#2}%
  \else% backwards compatibility: try \docutilsrole#1{#2}
    \ifcsname docutilsrole#1\endcsname%
      \csname docutilsrole#1\endcsname{#2}%
    \else%
      #2%
    \fi%
  \fi%
}

\ifthenelse{\isundefined{\hypersetup}}{
  \usepackage[colorlinks=true,linkcolor=blue,urlcolor=blue]{hyperref}
  \urlstyle{same} % normal text font (alternatives: tt, rm, sf)
}{}

\begin{document}
\newcounter{footnotecounter}\title{Optimised finite difference computation from symbolic equations}\author{Michael Lange$^{\setcounter{footnotecounter}{3}\fnsymbol{footnotecounter}\setcounter{footnotecounter}{1}\fnsymbol{footnotecounter}}$%
          \setcounter{footnotecounter}{1}\thanks{\fnsymbol{footnotecounter} %
          Corresponding author: \protect\href{mailto:michael.lange@imperial.ac.uk}{michael.lange@imperial.ac.uk}}\setcounter{footnotecounter}{3}\thanks{\fnsymbol{footnotecounter} Imperial College London}, Navjot Kukreja$^{\setcounter{footnotecounter}{3}\fnsymbol{footnotecounter}}$, Fabio Luporini$^{\setcounter{footnotecounter}{3}\fnsymbol{footnotecounter}}$, Mathias Louboutin$^{\setcounter{footnotecounter}{4}\fnsymbol{footnotecounter}}$\setcounter{footnotecounter}{4}\thanks{\fnsymbol{footnotecounter} The University of British Columbia}, Charles Yount$^{\setcounter{footnotecounter}{5}\fnsymbol{footnotecounter}}$\setcounter{footnotecounter}{5}\thanks{\fnsymbol{footnotecounter} Intel Corporation}, Jan Hückelheim$^{\setcounter{footnotecounter}{3}\fnsymbol{footnotecounter}}$, Gerard J. Gorman$^{\setcounter{footnotecounter}{3}\fnsymbol{footnotecounter}}$\thanks{%

          \noindent%
          Copyright\,\copyright\,2017 Michael Lange et al. This is an open-access article distributed under the terms of the Creative Commons Attribution License, which permits unrestricted use, distribution, and reproduction in any medium, provided the original author and source are credited.%
        }}\maketitle
          \renewcommand{\leftmark}{PROC. OF THE 15th PYTHON IN SCIENCE CONF. (SCIPY 2017)}
          \renewcommand{\rightmark}{OPTIMISED FINITE DIFFERENCE COMPUTATION FROM SYMBOLIC EQUATIONS}

\InputIfFileExists{page_numbers.tex}{}{}
\newcommand*{\docutilsroleref}{\ref}
\newcommand*{\docutilsrolelabel}{\label}
\providecommand*\DUrolecite[1]{\cite{#1}}
\begin{abstract}Domain-specific high-productivity environments are playing an
increasingly important role in scientific computing due to the
levels of abstraction and automation they provide. In this
paper we introduce Devito, an open-source domain-specific framework for
solving partial differential equations from symbolic problem
definitions by the finite difference method. We highlight the
generation and automated execution of highly optimized stencil code
from only a few lines of high-level symbolic Python for a set of
scientific equations, before exploring the use of Devito operators in
seismic inversion problems.\end{abstract}

\subsection{Introduction%
  \label{introduction}%
}

Domain-specific high-productivity environments are playing an
increasingly important role in scientific computing. The level of
abstraction and automation provided by such frameworks not only
increases productivity and accelerates innovation, but also allows the
combination of expertise from different specialised disciplines. This
synergy is necessary when creating the complex software stack needed
to solve leading edge scientific problems, since domain specialists as
well as high performance computing experts are required to fully
leverage modern computing architectures. Based on this philosophy we
introduce Devito \cite{Lange17}, an open-source domain-specific framework
for solving partial differential equations (PDE) from symbolic problem
definitions by the finite difference method.

Symbolic computation, where optimized numerical code is automatically
derived from a high-level problem definition, is a powerful technique
that allows domain scientists to focus on algorithmic development
rather than implementation details. For this reason Devito exposes an
API based on Python (SymPy) \cite{Meurer17} that allow users to express
equations symbolically, from which it generates and executes optimized
stencil code via just-in-time (JIT) compilation. Using latest advances
in stencil compiler research, Devito thus provides domain scientists
with the ability to quickly and efficiently generate high-performance
kernels from only a few lines of Python code, making Devito composable
with existing open-source software.

While Devito was originally developed for seismic imaging workflows,
the automated generation and optimization of stencil codes can be
utilised for a much broader set of computational problems. Matrix-free
stencil operators based on explicit finite difference schemes are
widely used in industry and academic research, although they merely
represent one of many approaches to solving PDEs \cite{Baba16}, \cite{Liu09},
\cite{Rai91}. In this paper we therefore limit our discussion of numerical
methods and instead focus on the ease with which these operators can be
created symbolically. We give a brief overview of the design concepts
and key features of Devito and demonstrate its API using a set of
classic examples from computational fluid dynamics (CFD). Then we will
discuss the use of Devito in an example of a complex seismic inversion
algorithm to illustrate its use in practical scientific applications
and to showcase the performance achieved by the auto-generated and
optimised code.

\subsection{Background%
  \label{background}%
}

The attraction of using domain-specific languages (DSL) to solve PDEs
via a high-level mathematical notation is by no means new and has led
to various purpose-built software packages and compilers dating back
to 1962 \cite{Iverson62}, \cite{Cardenas70,Umetani85,Cook88,VanEngelen96}. Following the emergence of Python as a widely used
programming language in scientific research, embedded DSLs for more
specialised domains came to the fore, most notably the FEniCS
\cite{Logg12} and Firedrake \cite{Rathgeber16} frameworks, which both implement
the unified Form Language (UFL) \cite{Alnaes14} for the symbolic
definition of finite element problems in the weak form. The increased
level of abstraction that such high-level languages provide decouples
the problem definition from its implementation, thus allowing domain
scientists and mathematicians to focus on more advanced methods, such
as the automation of adjoint models as demonstrated by Dolfin-Adjoint
\cite{Farrell13}.

The performance optimization of stencil computation on regular
cartesian grids for high-performance computing applications has also
received much attention in computer science research \cite{Datta08,Brandvik10,Zhang12,Henretty13,Yount15}. The primary focus
of most stencil compilers or DSLs, however, is the optimization of
synthetic problems which often limits their applicability for
practical scientific applications. The primary consideration here is
that most realistic problems often require more than just a fast and
efficient PDE solver, which entails that symbolic DSLs embedded in
Python can benefit greatly from native interoperability with the
scientific Python ecosystem.

\subsection{Design and API%
  \label{design-and-api}%
}

The primary objective of Devito is to enable the quick and effective
creation of highly optimised finite difference operators for use in a
realistic scientific application context. As such, its design is
centred around composability with the existing Python software stack
to provide users with the tools to dynamically generate optimised
stencil computation kernels and to enable access to the full
scientific software ecosystem. In addition, to accommodate the needs
of \textquotedbl{}real life\textquotedbl{} scientific applications, a secondary API is provided
that enables users to inject custom expressions, such as boundary
conditions or sparse point interpolation routines, into the generated
kernels.

The use of SymPy as the driver for the symbolic generation of stencil
expressions and the subsequent code-generation are at the heart of the
Devito philosophy. While SymPy is fully capable of auto-generating
low-level C code for pre-compiled execution from high-level symbolic
expressions, Devito is designed to combine these capabilities with
automatic performance optimization based on the latest advances in
stencil compiler technology. The result is a framework that is capable
of automatically generating and optimising complex stencil code from
high-level symbolic definitions.

The Devito API is based around two key concepts that allow users to
express finite difference problems in a concise symbolic notation:%
\begin{itemize}

\item 

\textbf{Symbolic data objects:} Devito's high-level symbolic objects
behave like \texttt{\DUrole{code}{sympy.Function}} objects and provide a set of
shorthand notations for generating derivative expressions, while
also managing user data. The rationale for this duality is that many
stencil optimization algorithms rely on data layout changes,
mandating that Devito needs to be in control of data allocation and
access.
\item 

\textbf{Operator:} An \texttt{\DUrole{code}{Operator}} creates, compiles and executes a
single executable kernel from a set of SymPy expressions. The code
generation and optimization process involves various stages and
accepts a mixture of high-level and low-level expressions to allow
the injection of customised code.
\end{itemize}

\subsection{Fluid Dynamics Examples%
  \label{fluid-dynamics-examples}%
}

In the following section we demonstrate the use of the Devito API to
implement two examples from classical fluid dynamics, before
highlighting the role of Devito operators in a seismic inversion
context. Both CFD examples are based in part on tutorials from the
introductory blog \textquotedbl{}CFD Python: 12 steps to Navier-Stokes\textquotedbl{}\DUfootnotemark{id20}{id22}{1} by the
Lorena A. Barba group. We have chosen the examples in this section for
their relative simplicity to concisely illustrate the capabilities
and API features of Devito. For a more complete discussion on
numerical methods for fluid flows please refer to \cite{Peiro05}.%
\DUfootnotetext{id22}{id20}{1}{
\url{http://lorenabarba.com/blog/cfd-python-12-steps-to-navier-stokes/}}

\subsubsection{Linear Convection%
  \label{linear-convection}%
}

We will demonstrate a basic Devito operator definition based on a
linear two-dimensional convection flow (step 5 in the original
tutorials)\DUfootnotemark{id23}{id24}{2}. The governing equation we are implementing here is:%
\DUfootnotetext{id24}{id23}{2}{
\url{http://nbviewer.jupyter.org/github/opesci/devito/blob/master/examples/cfd/test_01_convection_revisited.ipynb}}
\begin{equation}
\label{2dconvection}
\frac{\partial u}{\partial t}+c\frac{\partial u}{\partial x}
        + c\frac{\partial u}{\partial y} = 0
\end{equation}A discretised version of this equation, using a forward difference
scheme in time and a backward difference scheme in space might be written
as\begin{equation}
\label{2dconvdiscr}
u_{i,j}^{n+1} = u_{i,j}^n-c \frac{\Delta t}{\Delta x}(u_{i,j}^n-u_{i-1,j}^n)
- c \frac{\Delta t}{\Delta y}(u_{i,j}^n-u_{i,j-1}^n)
\end{equation}where the subscripts $i$ and $j$ denote indices in the
space dimensions and the superscript $n$ denotes the index in
time, while $\Delta t$, $\Delta x$, $\Delta y$
denote the spacing in time and space dimensions respectively.

The first thing we need is a function object with which we can build a
timestepping scheme. For this purpose Devito provides so-called
\texttt{\DUrole{code}{TimeData}} objects that encapsulate functions that are
differentiable in space and time. With this we can derive symbolic
expressions for the backward derivatives in space directly via the
\texttt{\DUrole{code}{u.dxl}} and \texttt{\DUrole{code}{u.dyl}} shorthand expressions (the \texttt{\DUrole{code}{l}}
indicates \textquotedbl{}left\textquotedbl{} or backward differences) and the shorthand notation
\texttt{\DUrole{code}{u.dt}} provided by \texttt{\DUrole{code}{TimeData}} objects to derive the
forward derivative in time.\vspace{1mm}
\begin{Verbatim}[commandchars=\\\{\},fontsize=\footnotesize]
\PY{k+kn}{from} \PY{n+nn}{devito} \PY{k+kn}{import} \PY{o}{*}

\PY{n}{c} \PY{o}{=} \PY{l+m+mf}{1.}
\PY{n}{u} \PY{o}{=} \PY{n}{TimeData}\PY{p}{(}\PY{n}{name}\PY{o}{=}\PY{l+s+s1}{\PYZsq{}}\PY{l+s+s1}{u}\PY{l+s+s1}{\PYZsq{}}\PY{p}{,} \PY{n}{shape}\PY{o}{=}\PY{p}{(}\PY{n}{nx}\PY{p}{,} \PY{n}{ny}\PY{p}{)}\PY{p}{)}

\PY{n}{eq} \PY{o}{=} \PY{n}{Eq}\PY{p}{(}\PY{n}{u}\PY{o}{.}\PY{n}{dt} \PY{o}{+} \PY{n}{c} \PY{o}{*} \PY{n}{u}\PY{o}{.}\PY{n}{dxl} \PY{o}{+} \PY{n}{c} \PY{o}{*} \PY{n}{u}\PY{o}{.}\PY{n}{dyl}\PY{p}{)}

\PY{p}{[}\PY{n}{In}\PY{p}{]} \PY{k}{print} \PY{n}{eq}
\PY{p}{[}\PY{n}{Out}\PY{p}{]} \PY{n}{Eq}\PY{p}{(}\PY{o}{\PYZhy{}}\PY{n}{u}\PY{p}{(}\PY{n}{t}\PY{p}{,} \PY{n}{x}\PY{p}{,} \PY{n}{y}\PY{p}{)}\PY{o}{/}\PY{n}{s} \PY{o}{+} \PY{n}{u}\PY{p}{(}\PY{n}{t} \PY{o}{+} \PY{n}{s}\PY{p}{,} \PY{n}{x}\PY{p}{,} \PY{n}{y}\PY{p}{)}\PY{o}{/}\PY{n}{s}
        \PY{o}{+} \PY{l+m+mf}{2.0}\PY{o}{*}\PY{n}{u}\PY{p}{(}\PY{n}{t}\PY{p}{,} \PY{n}{x}\PY{p}{,} \PY{n}{y}\PY{p}{)}\PY{o}{/}\PY{n}{h} \PY{o}{\PYZhy{}} \PY{l+m+mf}{1.0}\PY{o}{*}\PY{n}{u}\PY{p}{(}\PY{n}{t}\PY{p}{,} \PY{n}{x}\PY{p}{,} \PY{n}{y} \PY{o}{\PYZhy{}} \PY{n}{h}\PY{p}{)}\PY{o}{/}\PY{n}{h}
        \PY{o}{\PYZhy{}} \PY{l+m+mf}{1.0}\PY{o}{*}\PY{n}{u}\PY{p}{(}\PY{n}{t}\PY{p}{,} \PY{n}{x} \PY{o}{\PYZhy{}} \PY{n}{h}\PY{p}{,} \PY{n}{y}\PY{p}{)}\PY{o}{/}\PY{n}{h}\PY{p}{,} \PY{l+m+mi}{0}\PY{p}{)}
\end{Verbatim}
\vspace{1mm}
The above expression results in a \texttt{\DUrole{code}{sympy.Equation}} object that
contains the fully discretised form of Eq. \DUrole{ref}{2dconvection},
including placeholder symbols for grid spacing in space (\texttt{\DUrole{code}{h}},
assuming $\Delta x = \Delta y$) and time (\texttt{\DUrole{code}{s}}). These
spacing symbols will be resolved during the code generation process,
as described in the \hyperref[code-generation-section]{code generation section}. It is also important
to note here that the explicit generation of the space derivatives
\texttt{\DUrole{code}{u\_dx}} and \texttt{\DUrole{code}{u\_dy}} is due to the use of a backward
derivative in space to align with the original example. A similar
notation to the forward derivative in time (\texttt{\DUrole{code}{u.dt}}) will soon be
provided.

In order to create a functional \texttt{\DUrole{code}{Operator}} object, the
expression \texttt{\DUrole{code}{eq}} needs to be rearranged so that we may solve for
the unknown $u_{i,j}^{n+1}$. This is easily achieved by using
SymPy's \texttt{\DUrole{code}{solve}} utility and the Devito shorthand
\texttt{\DUrole{code}{u.forward}} which denotes the furthest forward stencil point in
a time derivative ($u_{i,j}^{n+1}$).\vspace{1mm}
\begin{Verbatim}[commandchars=\\\{\},fontsize=\footnotesize]
\PY{k+kn}{from} \PY{n+nn}{sympy} \PY{k+kn}{import} \PY{n}{solve}

\PY{n}{stencil} \PY{o}{=} \PY{n}{solve}\PY{p}{(}\PY{n}{eq}\PY{p}{,} \PY{n}{u}\PY{o}{.}\PY{n}{forward}\PY{p}{)}\PY{p}{[}\PY{l+m+mi}{0}\PY{p}{]}

\PY{p}{[}\PY{n}{In}\PY{p}{]} \PY{k}{print}\PY{p}{(}\PY{n}{stencil}\PY{p}{)}
\PY{p}{[}\PY{n}{Out}\PY{p}{]} \PY{p}{(}\PY{n}{h}\PY{o}{*}\PY{n}{u}\PY{p}{(}\PY{n}{t}\PY{p}{,} \PY{n}{x}\PY{p}{,} \PY{n}{y}\PY{p}{)} \PY{o}{\PYZhy{}} \PY{l+m+mf}{2.0}\PY{o}{*}\PY{n}{s}\PY{o}{*}\PY{n}{u}\PY{p}{(}\PY{n}{t}\PY{p}{,} \PY{n}{x}\PY{p}{,} \PY{n}{y}\PY{p}{)}
     \PY{o}{+} \PY{n}{s}\PY{o}{*}\PY{n}{u}\PY{p}{(}\PY{n}{t}\PY{p}{,} \PY{n}{x}\PY{p}{,} \PY{n}{y} \PY{o}{\PYZhy{}} \PY{n}{h}\PY{p}{)} \PY{o}{+} \PY{n}{s}\PY{o}{*}\PY{n}{u}\PY{p}{(}\PY{n}{t}\PY{p}{,} \PY{n}{x} \PY{o}{\PYZhy{}} \PY{n}{h}\PY{p}{,} \PY{n}{y}\PY{p}{)}\PY{p}{)}\PY{o}{/}\PY{n}{h}
\end{Verbatim}
\vspace{1mm}
The above variable \texttt{\DUrole{code}{stencil}} now represents the RHS of
Eq. \DUrole{ref}{2dconvdiscr}, allowing us to construct a SymPy expression
that updates $u_{i,j}^{n+1}$ and build a \texttt{\DUrole{code}{devito.Operator}}
from it. When creating this operator we also supply concrete values
for the spacing terms \texttt{\DUrole{code}{h}} and \texttt{\DUrole{code}{s}} via an additional
substitution map argument \texttt{\DUrole{code}{subs}}.\vspace{1mm}
\begin{Verbatim}[commandchars=\\\{\},fontsize=\footnotesize]
\PY{n}{op} \PY{o}{=} \PY{n}{Operator}\PY{p}{(}\PY{n}{Eq}\PY{p}{(}\PY{n}{u}\PY{o}{.}\PY{n}{forward}\PY{p}{,} \PY{n}{stencil}\PY{p}{)}\PY{p}{,}
              \PY{n}{subs}\PY{o}{=}\PY{p}{\PYZob{}}\PY{n}{h}\PY{p}{:} \PY{n}{dx}\PY{p}{,} \PY{n}{s}\PY{p}{:}\PY{n}{dt}\PY{p}{\PYZcb{}}\PY{p}{)}

\PY{c+c1}{\PYZsh{} Set initial condition as a smooth function}
\PY{n}{init\PYZus{}smooth}\PY{p}{(}\PY{n}{u}\PY{o}{.}\PY{n}{data}\PY{p}{,} \PY{n}{dx}\PY{p}{,} \PY{n}{dy}\PY{p}{)}

\PY{n}{op}\PY{p}{(}\PY{n}{u}\PY{o}{=}\PY{n}{u}\PY{p}{,} \PY{n}{time}\PY{o}{=}\PY{l+m+mi}{100}\PY{p}{)}  \PY{c+c1}{\PYZsh{} Apply for 100 timesteps}
\end{Verbatim}
\vspace{1mm}

\newpage
Using this operator we can now create a similar example to the one
presented in the original tutorial by initialising the data associated
with the symbolic function $u$, \texttt{\DUrole{code}{u.data}} with an initial
flow field. However, to avoid numerical errors due to the
discontinuities at the boundary of the original \textquotedbl{}hat function\textquotedbl{}, we use
the following smooth initial condition provided by \cite{Krakos12}, as
depicted in Figure \DUrole{ref}{fig2dconv}.\begin{equation*}
u_0(x,y)=1+u\left(\frac{2}{3}x\right)*u\left(\frac{2}{3}y\right)
\end{equation*}The final result after executing the operator for $5s$ (100
timesteps) is depicted in Figure \DUrole{ref}{fig2dconvfinal}. The result
shows the expected displacement of the initial shape, in accordance
with the prescribed velocity ($c = 1.0$), closely mirroring the
displacement of the \textquotedbl{}hat function\textquotedbl{} in the original tutorial. It should
also be noted that, while the results show good agreement with
expectations by visual inspection, they do not represent an accurate
solution to the linear convection equation. In particular, the low
order spatial discretisation introduces numerical diffusion that
causes a decrease in the peak velocity. This is a well-known issue
that could be addressed with more sophisticated solver schemes as
discussed in \cite{LeVeque92}.\begin{figure}[hbt]\noindent\makebox[\columnwidth][c]{\includegraphics[scale=0.42]{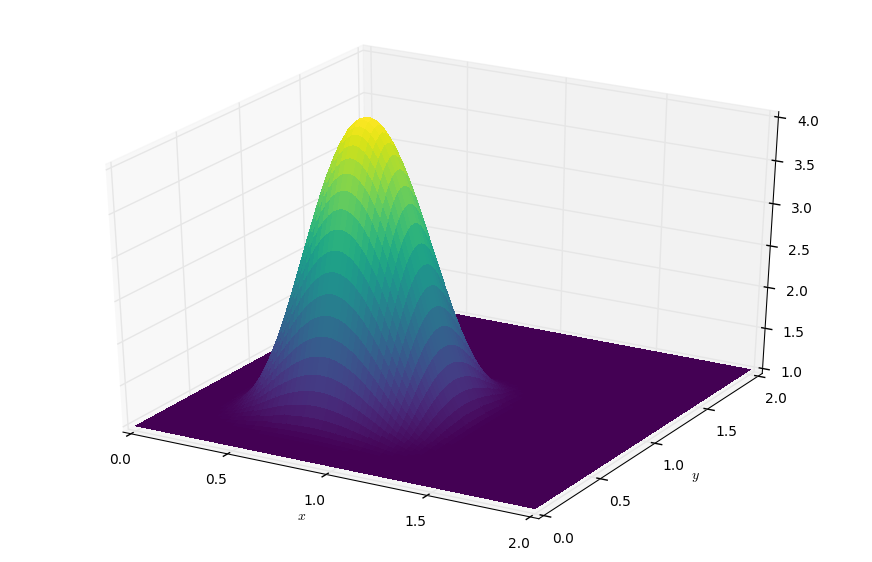}}
\caption{Initial condition of \texttt{\DUrole{code}{u.data}} in the 2D convection
example. \DUrole{label}{fig2dconv}}
\end{figure}\begin{figure}[hbt]\noindent\makebox[\columnwidth][c]{\includegraphics[scale=0.42]{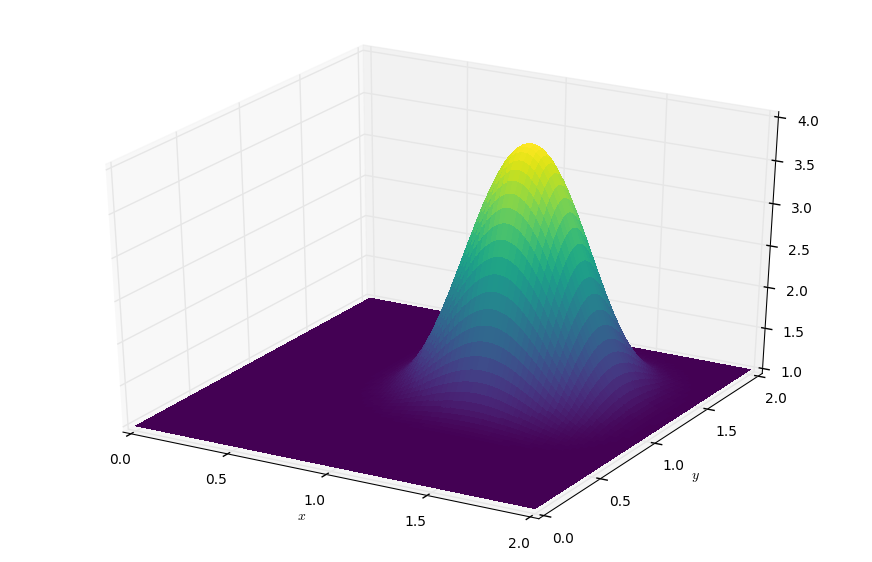}}
\caption{State of \texttt{\DUrole{code}{u.data}} after 100 timesteps in convection
example. \DUrole{label}{fig2dconvfinal}}
\end{figure}

\subsubsection{Laplace equation%
  \label{laplace-equation}%
}

The above example shows how Devito can be used to create finite
difference stencil operators from only a few lines of high-level
symbolic code. However, the previous example only required a single
variable to be updated, while more complex operators might need to
execute multiple expressions simultaneously, for example to solve
coupled PDEs or apply boundary conditions as part of the time
loop. For this reason \texttt{\DUrole{code}{devito.Operator}} objects can be
constructed from multiple update expressions and allow mutiple
expression formats as input.

Nevertheless, boundary conditions are currently not provided as part
of the symbolic high-level API. For exactly this reason,
Devito provides a low-level, or \textquotedbl{}indexed\textquotedbl{} API, where custom SymPy
expressions can be created with explicitly resolved grid accesses to
manually inject custom code into the auto-generation toolchain. This
entails that future extensions to capture different types of boundary
conditions can easily be added at a later stage.

To illustrate the use of the low-level API, we will use the Laplace
example from the original CFD tutorials (step 9), which implements the
steady-state heat equation with Dirichlet and Neuman boundary
conditions\DUfootnotemark{id27}{id28}{3}. The governing equation for this problem is%
\DUfootnotetext{id28}{id27}{3}{
\url{http://nbviewer.jupyter.org/github/opesci/devito/blob/master/examples/cfd/test_05_laplace.ipynb}}
\begin{equation}
\label{2dlaplace}
\frac{\partial ^2 p}{\partial x^2} + \frac{\partial ^2 p}{\partial y^2} = 0
\end{equation}The rearranged discretised form, assuming a central difference scheme
for second derivatives, is\begin{equation}
\label{2dlaplace_discr}
p_{i,j}^n = \frac{\Delta y^2(p_{i+1,j}^n+p_{i-1,j}^n)
        +\Delta x^2(p_{i,j+1}^n + p_{i,j-1}^n)}
        {2(\Delta x^2 + \Delta y^2)}
\end{equation}Using a similar approach to the previous example, we can construct
the SymPy expression to update the state of a field $p$. For
demonstration purposes we will use two separate function objects
of type \texttt{\DUrole{code}{DenseData}} in this example, since the Laplace equation
does not contain a time-dependence. The shorthand expressions
\texttt{\DUrole{code}{pn.dx2}} and \texttt{\DUrole{code}{pn.dy2}} hereby denote the second derivatives
in $x$ and $y$.\vspace{1mm}
\begin{Verbatim}[commandchars=\\\{\},fontsize=\footnotesize]
\PY{c+c1}{\PYZsh{} Create two separate symbols with space dimensions}
\PY{n}{p} \PY{o}{=} \PY{n}{DenseData}\PY{p}{(}\PY{n}{name}\PY{o}{=}\PY{l+s+s1}{\PYZsq{}}\PY{l+s+s1}{p}\PY{l+s+s1}{\PYZsq{}}\PY{p}{,} \PY{n}{shape}\PY{o}{=}\PY{p}{(}\PY{n}{nx}\PY{p}{,} \PY{n}{ny}\PY{p}{)}\PY{p}{,}
              \PY{n}{space\PYZus{}order}\PY{o}{=}\PY{l+m+mi}{2}\PY{p}{)}
\PY{n}{pn} \PY{o}{=} \PY{n}{DenseData}\PY{p}{(}\PY{n}{name}\PY{o}{=}\PY{l+s+s1}{\PYZsq{}}\PY{l+s+s1}{pn}\PY{l+s+s1}{\PYZsq{}}\PY{p}{,} \PY{n}{shape}\PY{o}{=}\PY{p}{(}\PY{n}{nx}\PY{p}{,} \PY{n}{ny}\PY{p}{)}\PY{p}{,}
               \PY{n}{space\PYZus{}order}\PY{o}{=}\PY{l+m+mi}{2}\PY{p}{)}

\PY{c+c1}{\PYZsh{} Define equation and solve for center point in `pn`}
\PY{n}{eq} \PY{o}{=} \PY{n}{Eq}\PY{p}{(}\PY{n}{a} \PY{o}{*} \PY{n}{pn}\PY{o}{.}\PY{n}{dx2} \PY{o}{+} \PY{n}{pn}\PY{o}{.}\PY{n}{dy2}\PY{p}{)}
\PY{n}{stencil} \PY{o}{=} \PY{n}{solve}\PY{p}{(}\PY{n}{eq}\PY{p}{,} \PY{n}{pn}\PY{p}{)}\PY{p}{[}\PY{l+m+mi}{0}\PY{p}{]}
\PY{c+c1}{\PYZsh{} The update expression to populate buffer `p`}
\PY{n}{eq\PYZus{}stencil} \PY{o}{=} \PY{n}{Eq}\PY{p}{(}\PY{n}{p}\PY{p}{,} \PY{n}{stencil}\PY{p}{)}
\end{Verbatim}
\vspace{1mm}
Just as the original tutorial, our initial condition in this example
is $p = 0$ and the flow will be driven by the boundary
conditions\begin{eqnarray*}
p=0\ &\text{at}\ x=0\\
p=y\ &\text{at}\ x=2\\
\frac{\partial p}{\partial y}=0\ &\text{at}\ y=0,\ 1
\end{eqnarray*}To implement these BCs we can utilise the \texttt{\DUrole{code}{.indexed}} property
that Devito symbols provide to get a symbol of type
\texttt{\DUrole{code}{sympy.IndexedBase}}, which in turn allows us to use matrix
indexing notation (square brackets) to create symbols of type
\texttt{\DUrole{code}{sympy.Indexed}} instead of \texttt{\DUrole{code}{sympy.Function}}. This notation
allows users to hand-code stencil expressions using explicit relative
grid indices, for example \texttt{\DUrole{code}{p{[}x, y{]} - p{[}x-1, y{]} / h}} for the
discretized backward derivative $\frac{\partial u}{\partial x}$.
The symbols \texttt{\DUrole{code}{x}} and \texttt{\DUrole{code}{y}} hereby represent the respective
problem dimensions and cause the expression to be executed over the
entire data dimension, similar to Python's \texttt{\DUrole{code}{:}} operator.

The Dirichlet BCs in the Laplace example can thus be implemented by
creating a \texttt{\DUrole{code}{sympy.Eq}} object that assigns either fixed values or
a prescribed function, such as the utility symbol \texttt{\DUrole{code}{bc\_right}} in
our example, along the left and right boundary of the domain. To
implement the Neumann BCs we again follow the original tutorial by
assigning the second grid row from the top and bottom boundaries the
value of the outermost row. The resulting SymPy expressions can then
be used alongside the state update expression to create our
\texttt{\DUrole{code}{Operator}} object.\vspace{1mm}
\begin{Verbatim}[commandchars=\\\{\},fontsize=\footnotesize]
\PY{c+c1}{\PYZsh{} Create an additional symbol for our prescibed BC}
\PY{n}{bc\PYZus{}right} \PY{o}{=} \PY{n}{DenseData}\PY{p}{(}\PY{n}{name}\PY{o}{=}\PY{l+s+s1}{\PYZsq{}}\PY{l+s+s1}{bc\PYZus{}right}\PY{l+s+s1}{\PYZsq{}}\PY{p}{,} \PY{n}{shape}\PY{o}{=}\PY{p}{(}\PY{n}{nx}\PY{p}{,} \PY{p}{)}\PY{p}{,}
                   \PY{n}{dimensions}\PY{o}{=}\PY{p}{(}\PY{n}{x}\PY{p}{,} \PY{p}{)}\PY{p}{)}
\PY{n}{bc\PYZus{}right}\PY{o}{.}\PY{n}{data}\PY{p}{[}\PY{p}{:}\PY{p}{]} \PY{o}{=} \PY{n}{np}\PY{o}{.}\PY{n}{linspace}\PY{p}{(}\PY{l+m+mi}{0}\PY{p}{,} \PY{l+m+mi}{1}\PY{p}{,} \PY{n}{nx}\PY{p}{)}

\PY{c+c1}{\PYZsh{} Create explicit boundary condition expressions}
\PY{n}{bc} \PY{o}{=} \PY{p}{[}\PY{n}{Eq}\PY{p}{(}\PY{n}{p}\PY{o}{.}\PY{n}{indexed}\PY{p}{[}\PY{n}{x}\PY{p}{,} \PY{l+m+mi}{0}\PY{p}{]}\PY{p}{,} \PY{l+m+mf}{0.}\PY{p}{)}\PY{p}{]}
\PY{n}{bc} \PY{o}{+}\PY{o}{=} \PY{p}{[}\PY{n}{Eq}\PY{p}{(}\PY{n}{p}\PY{o}{.}\PY{n}{indexed}\PY{p}{[}\PY{n}{x}\PY{p}{,} \PY{n}{ny}\PY{o}{\PYZhy{}}\PY{l+m+mi}{1}\PY{p}{]}\PY{p}{,} \PY{n}{bc\PYZus{}right}\PY{o}{.}\PY{n}{indexed}\PY{p}{[}\PY{n}{x}\PY{p}{]}\PY{p}{)}\PY{p}{]}
\PY{n}{bc} \PY{o}{+}\PY{o}{=} \PY{p}{[}\PY{n}{Eq}\PY{p}{(}\PY{n}{p}\PY{o}{.}\PY{n}{indexed}\PY{p}{[}\PY{l+m+mi}{0}\PY{p}{,} \PY{n}{y}\PY{p}{]}\PY{p}{,} \PY{n}{p}\PY{o}{.}\PY{n}{indexed}\PY{p}{[}\PY{l+m+mi}{1}\PY{p}{,} \PY{n}{y}\PY{p}{]}\PY{p}{)}\PY{p}{]}
\PY{n}{bc} \PY{o}{+}\PY{o}{=} \PY{p}{[}\PY{n}{Eq}\PY{p}{(}\PY{n}{p}\PY{o}{.}\PY{n}{indexed}\PY{p}{[}\PY{n}{nx}\PY{o}{\PYZhy{}}\PY{l+m+mi}{1}\PY{p}{,} \PY{n}{y}\PY{p}{]}\PY{p}{,} \PY{n}{p}\PY{o}{.}\PY{n}{indexed}\PY{p}{[}\PY{n}{nx}\PY{o}{\PYZhy{}}\PY{l+m+mi}{2}\PY{p}{,} \PY{n}{y}\PY{p}{]}\PY{p}{)}\PY{p}{]}

\PY{c+c1}{\PYZsh{} Build operator with update and BC expressions}
\PY{n}{op} \PY{o}{=} \PY{n}{Operator}\PY{p}{(}\PY{n}{expressions}\PY{o}{=}\PY{p}{[}\PY{n}{eq\PYZus{}stencil}\PY{p}{]} \PY{o}{+} \PY{n}{bc}\PY{p}{,}
              \PY{n}{subs}\PY{o}{=}\PY{p}{\PYZob{}}\PY{n}{h}\PY{p}{:} \PY{n}{dx}\PY{p}{,} \PY{n}{a}\PY{p}{:} \PY{l+m+mf}{1.}\PY{p}{\PYZcb{}}\PY{p}{)}
\end{Verbatim}
\vspace{1mm}
After building the operator, we can now use it in a time-independent
convergence loop that minimizes the $L^1$ norm of
$p$. However, in this example we need to make sure to explicitly
exchange the role of the buffers \texttt{\DUrole{code}{p}} and \texttt{\DUrole{code}{pn}}. This can
be achieved by supplying symbolic data objects via keyword arguments
when invoking the operator, where the name of the argument is matched
against the name of the original symbol used to create the operator.

The convergence criterion for this example is defined as the relative
error between two iterations and set to $\Vert p \Vert ^{1} <
10^{-4}$. The corresponding initial condition and the resulting
steady-state solution, depicted in Figures \DUrole{ref}{fig2dlaplace} and
\DUrole{ref}{fig2dlaplacefinal} respectively, agree with the original
tutorial implementation. It should again be noted that the chosen
numerical scheme might not be optimal to solve steady-state problems
of this type, since implicit methods are often preferred.\vspace{1mm}
\begin{Verbatim}[commandchars=\\\{\},fontsize=\footnotesize]
\PY{n}{l1norm} \PY{o}{=} \PY{l+m+mi}{1}
\PY{n}{counter} \PY{o}{=} \PY{l+m+mi}{0}
\PY{k}{while} \PY{n}{l1norm} \PY{o}{\PYZgt{}} \PY{l+m+mf}{1.e\PYZhy{}4}\PY{p}{:}
    \PY{c+c1}{\PYZsh{} Determine buffer order}
    \PY{k}{if} \PY{n}{counter} \PY{o}{\PYZpc{}} \PY{l+m+mi}{2} \PY{o}{==} \PY{l+m+mi}{0}\PY{p}{:}
        \PY{n}{\PYZus{}p}\PY{p}{,} \PY{n}{\PYZus{}pn} \PY{o}{=} \PY{n}{p}\PY{p}{,} \PY{n}{pn}
    \PY{k}{else}\PY{p}{:}
        \PY{n}{\PYZus{}p}\PY{p}{,} \PY{n}{\PYZus{}pn} \PY{o}{=} \PY{n}{pn}\PY{p}{,} \PY{n}{p}

    \PY{c+c1}{\PYZsh{} Apply operator}
    \PY{n}{op}\PY{p}{(}\PY{n}{p}\PY{o}{=}\PY{n}{\PYZus{}p}\PY{p}{,} \PY{n}{pn}\PY{o}{=}\PY{n}{\PYZus{}pn}\PY{p}{)}

    \PY{c+c1}{\PYZsh{} Compute L1 norm}
    \PY{n}{l1norm} \PY{o}{=} \PY{p}{(}\PY{n}{np}\PY{o}{.}\PY{n}{sum}\PY{p}{(}\PY{n}{np}\PY{o}{.}\PY{n}{abs}\PY{p}{(}\PY{n}{\PYZus{}p}\PY{o}{.}\PY{n}{data}\PY{p}{[}\PY{p}{:}\PY{p}{]}\PY{p}{)}
              \PY{o}{\PYZhy{}} \PY{n}{np}\PY{o}{.}\PY{n}{abs}\PY{p}{(}\PY{n}{\PYZus{}pn}\PY{o}{.}\PY{n}{data}\PY{p}{[}\PY{p}{:}\PY{p}{]}\PY{p}{)}\PY{p}{)}
              \PY{o}{/} \PY{n}{np}\PY{o}{.}\PY{n}{sum}\PY{p}{(}\PY{n}{np}\PY{o}{.}\PY{n}{abs}\PY{p}{(}\PY{n}{\PYZus{}pn}\PY{o}{.}\PY{n}{data}\PY{p}{[}\PY{p}{:}\PY{p}{]}\PY{p}{)}\PY{p}{)}\PY{p}{)}
    \PY{n}{counter} \PY{o}{+}\PY{o}{=} \PY{l+m+mi}{1}
\end{Verbatim}
\vspace{1mm}
\begin{figure}[]\noindent\makebox[\columnwidth][c]{\includegraphics[scale=0.42]{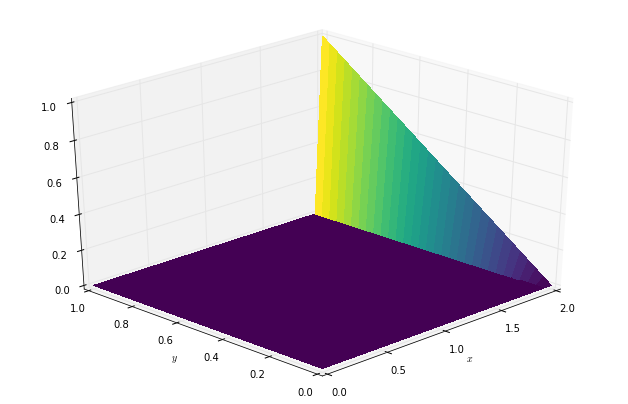}}
\caption{Initial condition of \texttt{\DUrole{code}{pn.data}} in the 2D Laplace
example. \DUrole{label}{fig2dlaplace}}
\end{figure}\begin{figure}[]\noindent\makebox[\columnwidth][c]{\includegraphics[scale=0.42]{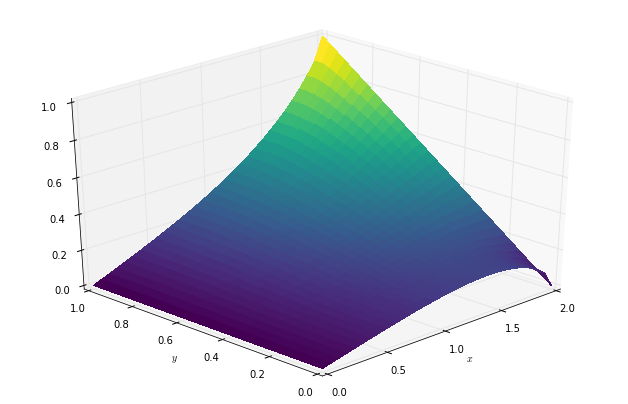}}
\caption{State of \texttt{\DUrole{code}{p.data}} after convergence in Laplace
example. \DUrole{label}{fig2dlaplacefinal}}
\end{figure}

\subsection{Seismic Inversion Example%
  \label{seismic-inversion-example}%
}
The primary motivating application behind the design of Devito is
the solution of seismic exploration problems that require highly
optimised wave propagation operators for forward modelling and
adjoint-based inversion. Obviously, the speed and accuracy of the
generated kernels are of vital importance. Moreover, the ability to
efficiently define rigorous forward modelling and adjoint operators
from high-level symbolic definitions also implies that domain
scientists are able to quickly adjust the numerical method and
discretisation to the individual problem and hardware architecture
\cite{Louboutin17a}.

In the following example  we will show the generation of forward and
adjoint operators for the acoustic wave equation and verify their
correctness using the so-called \emph{adjoint test} \cite{Virieux09}\DUfootnotemark{id31}{id32}{4}. This
test, also known as \emph{dot product test}, verifies that the
implementation of an adjoint operator indeed computes the conjugate
transpose of the forward operator.%
\DUfootnotetext{id32}{id31}{4}{
\url{http://nbviewer.jupyter.org/github/opesci/devito/blob/master/examples/seismic/tutorials/test_01_modelling.ipynb}}

The governing wave equation for the forward operator is defined as\begin{equation*}
m \frac{\partial^2 u}{\partial t^2}
+ \eta \frac{\partial u}{\partial t} - \nabla^2 u = q
\end{equation*}where $u$ denotes the pressure wave field, $m$ is the
square slowness, $q$ is the source term and $\eta$ denotes
the spatially varying dampening factor used to implement an absorbing
boundary condition.

On top of fast stencil operators, seismic inversion kernels also rely
on sparse point interpolation to inject the modelled wave as a point
source ($q$) and to record the pressure at individual point
locations. To accommodate this, Devito provides another symbolic data
type \texttt{\DUrole{code}{PointData}}, which allows the generation of sparse-point
interpolation expressions using the \textquotedbl{}indexed\textquotedbl{} low-level API. These
symbolic objects provide utility routines
\texttt{\DUrole{code}{pt.interpolate(expression)}} and \texttt{\DUrole{code}{pt.inject(field,
expression)}} to create symbolic expressions that perform linear
interpolation between the sparse points and the cartesian grid for
insertion into \texttt{\DUrole{code}{Operator}} kernels. A separate set of explicit
coordinate values is associated with the sparse point objects for
this purpose in addition to the function values stored in the
\texttt{\DUrole{code}{data}} property.

\subsubsection{Adjoint Test%
  \label{adjoint-test}%
}

The first step for implementing the adjoint test is to build a forward
operator that models the wave propagating through an isotropic medium,
where the square slowness of the wave is denoted as $m$.  Since
\texttt{\DUrole{code}{m}}, as well as the boundary dampening function \texttt{\DUrole{code}{eta}}, is
re-used between forward and adjoint runs the only symbolic data object
we need to create here is the wavefield \texttt{\DUrole{code}{u}} in order to
implement and rearrange our discretised equation \texttt{\DUrole{code}{eqn}} to form
the update expression for \texttt{\DUrole{code}{u}}. It is worth noting that the
\texttt{\DUrole{code}{u.laplace}} shorthand notation used here expands to the set of
second derivatives in all spatial dimensions, thus allowing us to use
the same formulation for two-dimensional and three-dimensional
problems.

In addition to the state update of \texttt{\DUrole{code}{u}}, we are also inserting
two additional terms into the forward modelling operator:%
\begin{itemize}

\item 

\texttt{\DUrole{code}{src\_term}} injects a pressure source at a point location
according to a prescribed time series stored in \texttt{\DUrole{code}{src.data}}
that is accessible in symbolic form via the symbol \texttt{\DUrole{code}{src}}.
The scaling factor in \texttt{\DUrole{code}{src\_term}} is coded by hand but can
be automatically inferred.
\item 

\texttt{\DUrole{code}{rec\_term}} adds the expression to interpolate the wavefield
\texttt{\DUrole{code}{u}} for a set of \textquotedbl{}receiver\textquotedbl{} hydrophones that measure the
propagated wave at varying distances from the source for every time
step. The resulting interpolated point data will be stored in
\texttt{\DUrole{code}{rec.data}} and is accessible to the user as a NumPy array.
\end{itemize}
\vspace{1mm}
\begin{Verbatim}[commandchars=\\\{\},fontsize=\footnotesize]
\PY{k}{def} \PY{n+nf}{forward}\PY{p}{(}\PY{n}{model}\PY{p}{,} \PY{n}{m}\PY{p}{,} \PY{n}{eta}\PY{p}{,} \PY{n}{src}\PY{p}{,} \PY{n}{rec}\PY{p}{,} \PY{n}{order}\PY{o}{=}\PY{l+m+mi}{2}\PY{p}{)}\PY{p}{:}
    \PY{c+c1}{\PYZsh{} Create the wavefeld function}
    \PY{n}{u} \PY{o}{=} \PY{n}{TimeData}\PY{p}{(}\PY{n}{name}\PY{o}{=}\PY{l+s+s1}{\PYZsq{}}\PY{l+s+s1}{u}\PY{l+s+s1}{\PYZsq{}}\PY{p}{,} \PY{n}{shape}\PY{o}{=}\PY{n}{model}\PY{o}{.}\PY{n}{shape}\PY{p}{,}
                 \PY{n}{time\PYZus{}order}\PY{o}{=}\PY{l+m+mi}{2}\PY{p}{,} \PY{n}{space\PYZus{}order}\PY{o}{=}\PY{n}{order}\PY{p}{)}

    \PY{c+c1}{\PYZsh{} Derive stencil from symbolic equation}
    \PY{n}{eqn} \PY{o}{=} \PY{n}{m} \PY{o}{*} \PY{n}{u}\PY{o}{.}\PY{n}{dt2} \PY{o}{\PYZhy{}} \PY{n}{u}\PY{o}{.}\PY{n}{laplace} \PY{o}{+} \PY{n}{eta} \PY{o}{*} \PY{n}{u}\PY{o}{.}\PY{n}{dt}
    \PY{n}{stencil} \PY{o}{=} \PY{n}{solve}\PY{p}{(}\PY{n}{eqn}\PY{p}{,} \PY{n}{u}\PY{o}{.}\PY{n}{forward}\PY{p}{)}\PY{p}{[}\PY{l+m+mi}{0}\PY{p}{]}
    \PY{n}{update\PYZus{}u} \PY{o}{=} \PY{p}{[}\PY{n}{Eq}\PY{p}{(}\PY{n}{u}\PY{o}{.}\PY{n}{forward}\PY{p}{,} \PY{n}{stencil}\PY{p}{)}\PY{p}{]}


    \PY{c+c1}{\PYZsh{} Add source injection and receiver interpolation}
    \PY{n}{src\PYZus{}term} \PY{o}{=} \PY{n}{src}\PY{o}{.}\PY{n}{inject}\PY{p}{(}\PY{n}{field}\PY{o}{=}\PY{n}{u}\PY{p}{,}
                          \PY{n}{expr}\PY{o}{=}\PY{n}{src} \PY{o}{*} \PY{n}{dt}\PY{o}{*}\PY{o}{*}\PY{l+m+mi}{2} \PY{o}{/} \PY{n}{m}\PY{p}{)}
    \PY{n}{rec\PYZus{}term} \PY{o}{=} \PY{n}{rec}\PY{o}{.}\PY{n}{interpolate}\PY{p}{(}\PY{n}{expr}\PY{o}{=}\PY{n}{u}\PY{p}{)}

    \PY{c+c1}{\PYZsh{} Create operator with source and receiver terms}
    \PY{k}{return} \PY{n}{Operator}\PY{p}{(}\PY{n}{update\PYZus{}u} \PY{o}{+} \PY{n}{src\PYZus{}term} \PY{o}{+} \PY{n}{rec\PYZus{}term}\PY{p}{,}
                    \PY{n}{subs}\PY{o}{=}\PY{p}{\PYZob{}}\PY{n}{s}\PY{p}{:} \PY{n}{dt}\PY{p}{,} \PY{n}{h}\PY{p}{:} \PY{n}{model}\PY{o}{.}\PY{n}{spacing}\PY{p}{\PYZcb{}}\PY{p}{)}
\end{Verbatim}
\vspace{1mm}
After building a forward operator, we can now implement the adjoint
operator in a similar fashion. Using the provided symbols \texttt{\DUrole{code}{m}}
and \texttt{\DUrole{code}{eta}}, we can again define the adjoint wavefield \texttt{\DUrole{code}{v}}
and implement its update expression from the discretised
equation. However, since the adjoint operator needs to operate
backwards in time there are two notable differences:%
\begin{itemize}

\item 

The update expression now updates the backward stencil point in the
time derivative $v_{i,j}^{n-1}$, denoted as
\texttt{\DUrole{code}{v.backward}}.  In addition to that, the \texttt{\DUrole{code}{Operator}} is
forced to reverse its internal time loop by providing the argument
\texttt{\DUrole{code}{time\_axis=Backward}}
\item 

Since the acoustic wave equation is self-adjoint without dampening,
the only change required in the governing equation is to invert the
sign of the dampening term \texttt{\DUrole{code}{eta * u.dt}}. The first derivative
is an antisymmetric operator and its adjoint minus itself.
\end{itemize}

Moreover, the role of the sparse point objects has now switched:
Instead of injecting the source term, we are now injecting the
previously recorded receiver values into the adjoint wavefield, while
we are interpolating the resulting wave at the original source
location. As the injection and interpolations are part of the kernel,
we also insure that these two are adjoints of each other.\vspace{1mm}
\begin{Verbatim}[commandchars=\\\{\},fontsize=\footnotesize]
\PY{k}{def} \PY{n+nf}{adjoint}\PY{p}{(}\PY{n}{model}\PY{p}{,} \PY{n}{m}\PY{p}{,} \PY{n}{eta}\PY{p}{,} \PY{n}{srca}\PY{p}{,} \PY{n}{rec}\PY{p}{,} \PY{n}{order}\PY{o}{=}\PY{l+m+mi}{2}\PY{p}{)}\PY{p}{:}
    \PY{c+c1}{\PYZsh{} Create the adjoint wavefeld function}
    \PY{n}{v} \PY{o}{=} \PY{n}{TimeData}\PY{p}{(}\PY{n}{name}\PY{o}{=}\PY{l+s+s1}{\PYZsq{}}\PY{l+s+s1}{v}\PY{l+s+s1}{\PYZsq{}}\PY{p}{,} \PY{n}{shape}\PY{o}{=}\PY{n}{model}\PY{o}{.}\PY{n}{shape}\PY{p}{,}
                 \PY{n}{time\PYZus{}order}\PY{o}{=}\PY{l+m+mi}{2}\PY{p}{,} \PY{n}{space\PYZus{}order}\PY{o}{=}\PY{n}{order}\PY{p}{)}

    \PY{c+c1}{\PYZsh{} Derive stencil from symbolic equation}
    \PY{c+c1}{\PYZsh{} Note the inversion of the dampening term}
    \PY{n}{eqn} \PY{o}{=} \PY{n}{m} \PY{o}{*} \PY{n}{v}\PY{o}{.}\PY{n}{dt2} \PY{o}{\PYZhy{}} \PY{n}{v}\PY{o}{.}\PY{n}{laplace} \PY{o}{\PYZhy{}} \PY{n}{eta} \PY{o}{*} \PY{n}{v}\PY{o}{.}\PY{n}{dt}
    \PY{n}{stencil} \PY{o}{=} \PY{n}{solve}\PY{p}{(}\PY{n}{eqn}\PY{p}{,} \PY{n}{u}\PY{o}{.}\PY{n}{forward}\PY{p}{)}\PY{p}{[}\PY{l+m+mi}{0}\PY{p}{]}
    \PY{n}{update\PYZus{}v} \PY{o}{=} \PY{p}{[}\PY{n}{Eq}\PY{p}{(}\PY{n}{v}\PY{o}{.}\PY{n}{backward}\PY{p}{,} \PY{n}{stencil}\PY{p}{)}\PY{p}{]}

    \PY{c+c1}{\PYZsh{} Inject the previous receiver readings}
    \PY{n}{rec\PYZus{}term} \PY{o}{=} \PY{n}{rec}\PY{o}{.}\PY{n}{inject}\PY{p}{(}\PY{n}{field}\PY{o}{=}\PY{n}{v}\PY{p}{,}
                          \PY{n}{expr}\PY{o}{=}\PY{n}{rec} \PY{o}{*} \PY{n}{dt}\PY{o}{*}\PY{o}{*}\PY{l+m+mi}{2} \PY{o}{/} \PY{n}{m}\PY{p}{)}

    \PY{c+c1}{\PYZsh{} Interpolate the adjoint\PYZhy{}source}
    \PY{n}{srca\PYZus{}term} \PY{o}{=} \PY{n}{srca}\PY{o}{.}\PY{n}{interpolate}\PY{p}{(}\PY{n}{expr}\PY{o}{=}\PY{n}{v}\PY{p}{)}

    \PY{c+c1}{\PYZsh{} Create operator with source and receiver terms}
    \PY{k}{return} \PY{n}{Operator}\PY{p}{(}\PY{n}{update\PYZus{}v} \PY{o}{+} \PY{n}{rec\PYZus{}term} \PY{o}{+} \PY{n}{srca\PYZus{}term}\PY{p}{,}
                    \PY{n}{subs}\PY{o}{=}\PY{p}{\PYZob{}}\PY{n}{s}\PY{p}{:} \PY{n}{dt}\PY{p}{,} \PY{n}{h}\PY{p}{:} \PY{n}{model}\PY{o}{.}\PY{n}{spacing}\PY{p}{\PYZcb{}}\PY{p}{,}
                    \PY{n}{time\PYZus{}axis}\PY{o}{=}\PY{n}{Backward}\PY{p}{)}
\end{Verbatim}
\vspace{1mm}
Having established how to build the required operators we can now
define the workflow for our adjoint example.  For illustration
purposes we are using a utility object \texttt{\DUrole{code}{Model}} that provides the
core information for seismic inversion runs, such as the values for
\texttt{\DUrole{code}{m}} and the dampening term \texttt{\DUrole{code}{eta}}, as well as the
coordinates of the point source and receiver hydrophones. It is worth
noting that the spatial discretisation and thus the stencil size of
the operators is still fully parameterisable.\vspace{1mm}
\begin{Verbatim}[commandchars=\\\{\},fontsize=\footnotesize]
\PY{c+c1}{\PYZsh{} Create the seismic model of the domain}
\PY{n}{model} \PY{o}{=} \PY{n}{Model}\PY{p}{(}\PY{o}{.}\PY{o}{.}\PY{o}{.}\PY{p}{)}

\PY{c+c1}{\PYZsh{} Create source with Ricker wavelet}
\PY{n}{src} \PY{o}{=} \PY{n}{PointData}\PY{p}{(}\PY{n}{name}\PY{o}{=}\PY{l+s+s1}{\PYZsq{}}\PY{l+s+s1}{src}\PY{l+s+s1}{\PYZsq{}}\PY{p}{,} \PY{n}{ntime}\PY{o}{=}\PY{n}{ntime}\PY{p}{,}
                \PY{n}{ndim}\PY{o}{=}\PY{l+m+mi}{2}\PY{p}{,} \PY{n}{npoint}\PY{o}{=}\PY{l+m+mi}{1}\PY{p}{)}
\PY{n}{src}\PY{o}{.}\PY{n}{data}\PY{p}{[}\PY{l+m+mi}{0}\PY{p}{,} \PY{p}{:}\PY{p}{]} \PY{o}{=} \PY{n}{ricker\PYZus{}wavelet}\PY{p}{(}\PY{n}{ntime}\PY{p}{)}
\PY{n}{src}\PY{o}{.}\PY{n}{coordinates}\PY{o}{.}\PY{n}{data}\PY{p}{[}\PY{p}{:}\PY{p}{]} \PY{o}{=} \PY{n}{source\PYZus{}coords}

\PY{c+c1}{\PYZsh{} Create empty set of receivers}
\PY{n}{rec} \PY{o}{=} \PY{n}{PointData}\PY{p}{(}\PY{n}{name}\PY{o}{=}\PY{l+s+s1}{\PYZsq{}}\PY{l+s+s1}{rec}\PY{l+s+s1}{\PYZsq{}}\PY{p}{,} \PY{n}{ntime}\PY{o}{=}\PY{n}{ntime}\PY{p}{,}
                \PY{n}{ndim}\PY{o}{=}\PY{l+m+mi}{2}\PY{p}{,} \PY{n}{npoint}\PY{o}{=}\PY{l+m+mi}{101}\PY{p}{)}
\PY{n}{rec}\PY{o}{.}\PY{n}{coordinates}\PY{o}{.}\PY{n}{data}\PY{p}{[}\PY{p}{:}\PY{p}{]} \PY{o}{=} \PY{n}{receiver\PYZus{}coords}

\PY{c+c1}{\PYZsh{} Create empty adjoint source symbol}
\PY{n}{srca} \PY{o}{=} \PY{n}{PointData}\PY{p}{(}\PY{n}{name}\PY{o}{=}\PY{l+s+s1}{\PYZsq{}}\PY{l+s+s1}{srca}\PY{l+s+s1}{\PYZsq{}}\PY{p}{,} \PY{n}{ntime}\PY{o}{=}\PY{n}{ntime}\PY{p}{,}
                 \PY{n}{ndim}\PY{o}{=}\PY{l+m+mi}{2}\PY{p}{,} \PY{n}{npoint}\PY{o}{=}\PY{l+m+mi}{1}\PY{p}{)}
\PY{n}{srca}\PY{o}{.}\PY{n}{coordinates}\PY{o}{.}\PY{n}{data}\PY{p}{[}\PY{p}{:}\PY{p}{]} \PY{o}{=} \PY{n}{source\PYZus{}coords}

\PY{c+c1}{\PYZsh{} Create symbol for square slowness}
\PY{n}{m} \PY{o}{=} \PY{n}{DenseData}\PY{p}{(}\PY{n}{name}\PY{o}{=}\PY{l+s+s1}{\PYZsq{}}\PY{l+s+s1}{m}\PY{l+s+s1}{\PYZsq{}}\PY{p}{,} \PY{n}{shape}\PY{o}{=}\PY{n}{model}\PY{o}{.}\PY{n}{shape}\PY{p}{,}
              \PY{n}{space\PYZus{}order}\PY{o}{=}\PY{n}{order}\PY{p}{)}
\PY{n}{m}\PY{o}{.}\PY{n}{data}\PY{p}{[}\PY{p}{:}\PY{p}{]} \PY{o}{=} \PY{n}{model}  \PY{c+c1}{\PYZsh{} Set m from model data}



\PY{c+c1}{\PYZsh{} Create dampening term from model}
\PY{n}{eta} \PY{o}{=} \PY{n}{DenseData}\PY{p}{(}\PY{n}{name}\PY{o}{=}\PY{l+s+s1}{\PYZsq{}}\PY{l+s+s1}{eta}\PY{l+s+s1}{\PYZsq{}}\PY{p}{,} \PY{n}{shape}\PY{o}{=}\PY{n}{shape}\PY{p}{,}
                \PY{n}{space\PYZus{}order}\PY{o}{=}\PY{n}{order}\PY{p}{)}
\PY{n}{eta}\PY{o}{.}\PY{n}{data}\PY{p}{[}\PY{p}{:}\PY{p}{]} \PY{o}{=} \PY{n}{model}\PY{o}{.}\PY{n}{dampening}

\PY{c+c1}{\PYZsh{} Execute foward and adjoint runs}
\PY{n}{fwd} \PY{o}{=} \PY{n}{forward}\PY{p}{(}\PY{n}{model}\PY{p}{,} \PY{n}{m}\PY{p}{,} \PY{n}{eta}\PY{p}{,} \PY{n}{src}\PY{p}{,} \PY{n}{rec}\PY{p}{)}
\PY{n}{fwd}\PY{p}{(}\PY{n}{time}\PY{o}{=}\PY{n}{ntime}\PY{p}{)}
\PY{n}{adj} \PY{o}{=} \PY{n}{adjoint}\PY{p}{(}\PY{n}{model}\PY{p}{,} \PY{n}{m}\PY{p}{,} \PY{n}{eta}\PY{p}{,} \PY{n}{srca}\PY{p}{,} \PY{n}{rec}\PY{p}{)}
\PY{n}{adj}\PY{p}{(}\PY{n}{time}\PY{o}{=}\PY{n}{ntime}\PY{p}{)}

\PY{c+c1}{\PYZsh{} Test prescribed against adjoint source}
\PY{n}{adjoint\PYZus{}test}\PY{p}{(}\PY{n}{src}\PY{o}{.}\PY{n}{data}\PY{p}{,} \PY{n}{srca}\PY{o}{.}\PY{n}{data}\PY{p}{)}
\end{Verbatim}
\vspace{1mm}
The adjoint test is the core definition of the adjoint of a linear
operator. The mathematical correctness of the adjoint is required for
mathematical adjoint-based optimizations methods that are only
guarantied to converged with the correct adjoint. The test can be
written as:\begin{equation*}
<src,\ adjoint(rec)> = <forward(src),\ rec>
\end{equation*}The adjoint test can be used to verify the accuracy of the forward
propagation and adjoint operators and has been shown to agree for 2D
and 3D implementations \cite{Louboutin17b}. The shot record of the data
measured at the receiver locations after the forward run is shown in
Figure \DUrole{ref}{figshotrecord}.\begin{figure}[]\noindent\makebox[\columnwidth][c]{\includegraphics[scale=0.50]{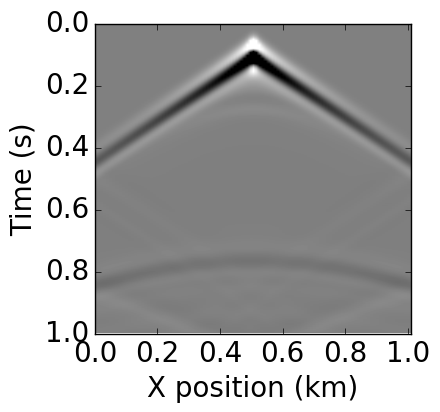}}
\caption{Shot record of the measured point values in \texttt{\DUrole{code}{rec.data}} after
the forward run. \DUrole{label}{figshotrecord}}
\end{figure}

\subsection{Automated code generation%
  \label{automated-code-generation}%
  \label{code-generation-section}%
}

The role of the \texttt{\DUrole{code}{Operator}} in the previous examples is to
generate semantically equivalent C code to the provided SymPy
expressions, complete with loop constructs and annotations for
performance optimization, such as OpenMP pragmas. Unlike many other
DSL-based frameworks, Devito employs actual compiler technology during
the code generation and optimization process. The symbolic
specification is progressively lowered to C code through a series of
passes manipulating abstract syntax trees (AST), rather than working
with rigid templates. This software engineering choice has an
invaluable impact on maintainability, extensibility and composability.

Following the initial resolution of explicit grid indices into the
low-level format, Devito is able to apply several types of automated
performance optimization throughout the code generation pipeline,
which are grouped into two distinct sub-modules:%
\begin{itemize}

\item 

\textbf{DSE - Devito Symbolic Engine:} The first set of optimization
passes consists of manipulating SymPy equations with the aim to
decrease the number of floating-point operations performed when
evaluating a single grid point. This initial optimization is
performed following an initial analysis of the provided expressions
and consists of sub-passes such as common sub-expressions
elimination, detection and promotion of time-invariants, and
factorization of common finite-difference weights. These
transformations not only optimize the operation count, but they also
improve the symbolic processing and low-level compilation times of
later processing stages.
\item 

\textbf{DLE - Devito Loop Engine:} After the initial symbolic processing
Devito schedules the optimised expressions in a set of loops by
creating an Abstract Syntax Tree (AST). The loop engine (DLE) is now
able to perform typical loop-level optimizations in mutiple passes
by manipulating this AST, including data alignment through array
annotations and padding, SIMD vectorization through OpenMP pragmas
and thread parallelism through OpenMP pragmas. On top of that, loop
blocking is used to fully exploit the memory bandwidth of a target
architecture by increasing data locality and thus cache
utilization. Since the effectiveness of the blocking technique is
highly architecture-dependent, Devito can determine optimal block
size through runtime auto-tuning.
\end{itemize}

\subsubsection{Performance Benchmark%
  \label{performance-benchmark}%
}

The effectiveness of the automated performance optimization performed
by the Devito backend engines can be demonstrated using the forward
operator constructed in the above example. The following performance
benchmarks were run with for a three-dimensional grid of size
$512\times512\times512$ with varying spatial discretisations
resulting in different stencil sizes with increasing operational
intensity (OI). The benchmark runs were performed on on a Intel(R)
Xeon E5-2620 v4 2.1Ghz \textquotedbl{}Broadwell\textquotedbl{} CPU with a single memory socket and
8 cores per socket and the slope of the roofline models was derived
using the Stream Triad benchmark \cite{McCalpin95}.

The first set of benchmark results, shown in Figure \DUrole{ref}{figperfdle},
highlights the performance gains achieved through loop-level
optimizations. For these runs the symbolic optimizations were kept at
a \textquotedbl{}basic\textquotedbl{} setting, where only common sub-expressions elimination is
performed on the kernel expressions. Of particular interest are the
performance gains achieved by increasing the loop engine mode from
\textquotedbl{}basic\textquotedbl{} to \textquotedbl{}advanced\textquotedbl{}, to insert loop blocking and explicit
vectorization directives into the generated code. Due to the improved
memory bandwidth utilization the performance increased to between
52\% and 74\% of the achievable peak. It is also worth noting that more
aggressive optimization in the \textquotedbl{}speculative\textquotedbl{} DLE mode (directives for
non-temporal stores and row-wise data alignment through additional
padding) did not yield any consistent improvements due to the low OI
inherent to the acoustic formulation of the wave equation and the
subsequent memory bandwidth limitations of the kernel.\begin{figure}[]\noindent\makebox[\columnwidth][c]{\includegraphics[scale=0.70]{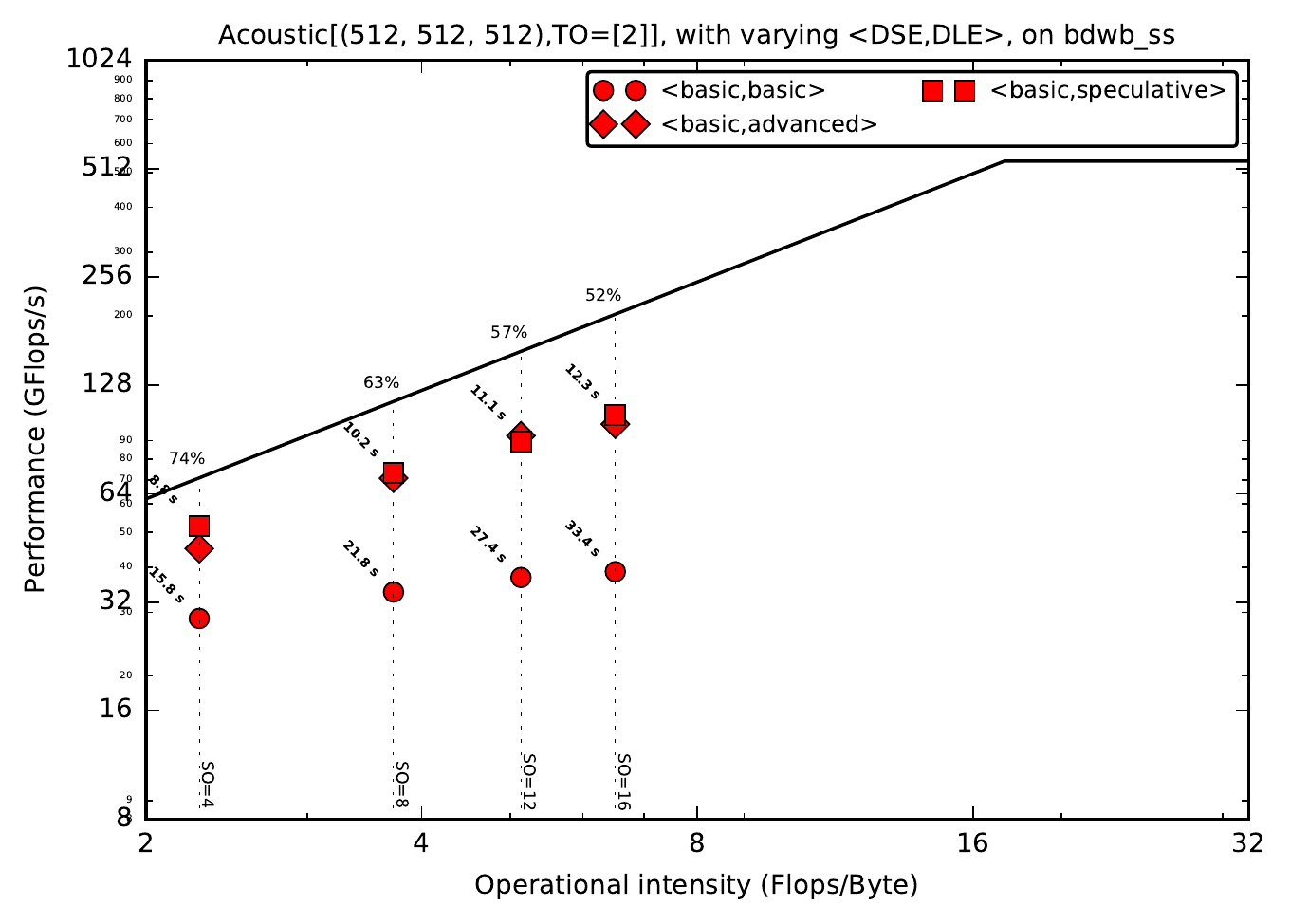}}
\caption{Performance benchmarks for loop-level optimizations with different
spatial orders (SO). The symbolic optimisations (DSE) have been
kept at level 'basic', while loop optimisation levels (DLE)
vary. \DUrole{label}{figperfdle}}
\end{figure}

On top of loop-level performance optimizations, Figure
\DUrole{ref}{figmaxperf} shows the achieved performance with additional
symbolic optimizations and flop reductions enabled. While the peak
performance shows only small effects from this set of optimizations
due to the inherent memory bandwidth limitations of the kernel, it is
interesting to note a reduction in operational intensity between
equivalent stencil sizes in Figures \DUrole{ref}{figperfdle} and
\DUrole{ref}{figmaxperf}. This entails that, despite only marginal runtime
changes, the generated code is performing less flops per stencil
point, which is of vital importance for compute-dominated kernels with
large OI \cite{Louboutin17a}.\begin{figure}[]\noindent\makebox[\columnwidth][c]{\includegraphics[scale=0.70]{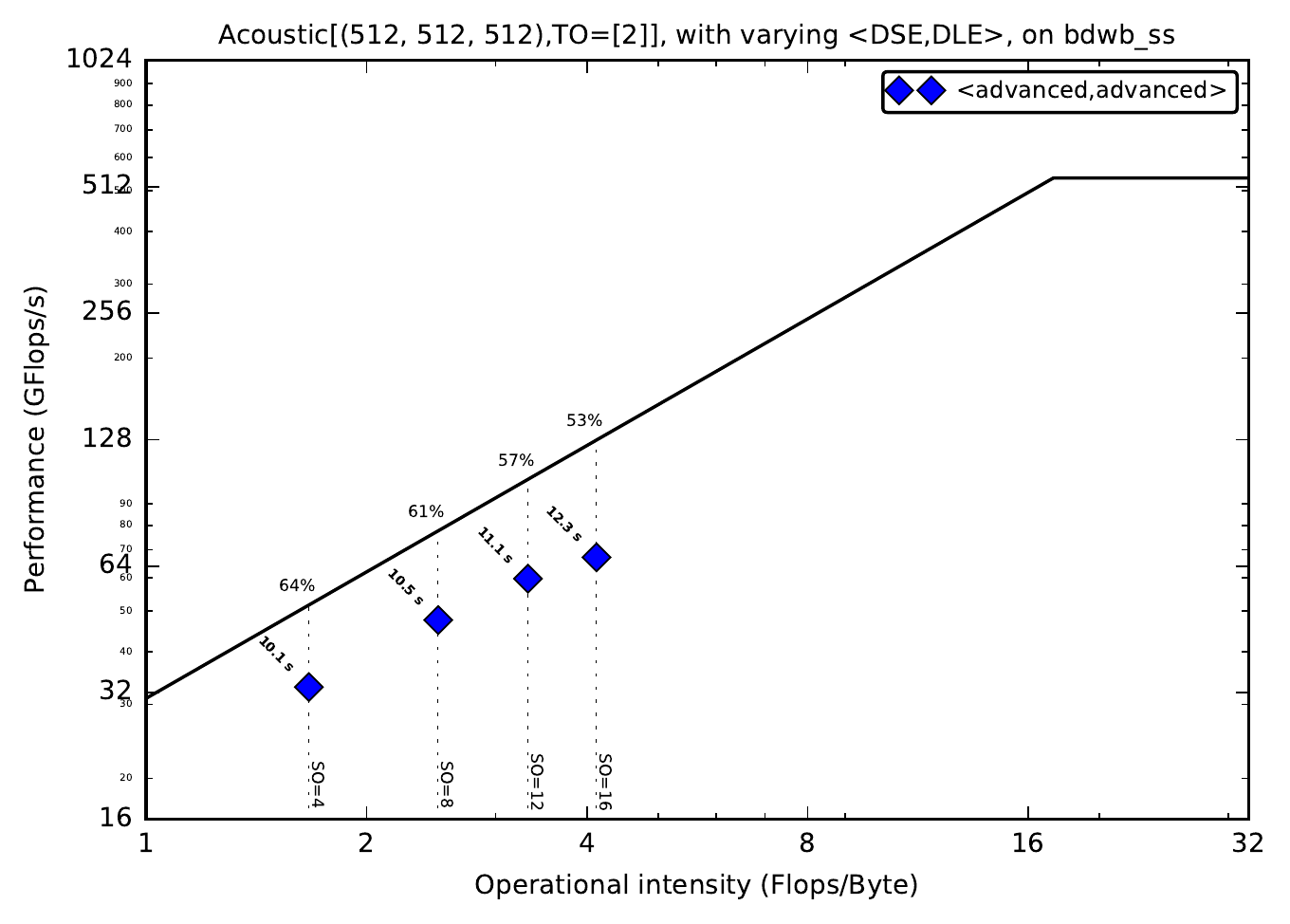}}
\caption{Performance benchmarks with full symbolic and loop-level
optimizations for different spatial orders
(SO). \DUrole{label}{figmaxperf}}
\end{figure}

\subsubsection{Integration with YASK%
  \label{integration-with-yask}%
}

As mentioned previously, Devito is based upon actual compiler
technology with a highly modular structure. Each backend
transformation pass is based on manipulating an input AST and
returning a new, different AST. One of the reasons behind this
software engineering strategy, which is clearly more challenging than
a template-based solution, is to ease the integration of external
tools, such as the YASK stencil optimizer \cite{Yount16}. We are currently
in the process of integrating YASK to complement the DLE, so that YASK
may replace some (but not all) DLE passes.

The DLE passes are organized in a hierarchy of classes where each
class represents a specific code transformation pipeline based on AST
manipulations. Integrating YASK becomes then a conceptually simple
task, which boils down to three actions:\newcounter{listcnt0}
\begin{list}{\arabic{listcnt0}.}
{
\usecounter{listcnt0}
\setlength{\rightmargin}{\leftmargin}
}

\item 

Adding a new transformation pipeline to the DLE.
\item 

Adding a new array type, to ease storage layout transformations
and data views (YASK employs a data layout different than the
conventional row-major format).
\item 

Creating the proper Python bindings in YASK so that Devito can
drive the code generation process.\end{list}

It has been shown that real-world stencil codes optimised through YASK
may achieve an exceptionally high fraction of the attainable machine
peak \cite{Yount15,Yount16}.  Further, initial prototyping (manual
optimization of Devito-generated code through YASK) revealed that YASK
may also outperform the loop optimization engine currently available
in Devito, besides ensuring seamless performance portability across a
range of computer architectures. On the other hand, YASK is a C++
based framework that, unlike Devito, does not rely on symbolic
mathematics and processing; in other words, it operates at a much
lower level of abstraction. These observations, as well as the outcome
of the initial prototyping phase, motivate the on-going Devito-YASK
integration effort.

\subsection{Discussion%
  \label{discussion}%
}

In this paper we present the finite difference DSL Devito and
demonstrate its high-level API to generate two fluid dynamics
operators and a full seismic inversion example. We highlight the
relative ease with which to create complex operators from only a few
lines of high-level Python code while utilising highly optimised
auto-generated C kernels via JIT compilation. On top of purely
symbolic top-level API based on SymPy, we show how to utilise Devito's
secondary API to inject custom expressions into the code generation
toolchain to implement Dirichlet and Neumann boundary conditions, as
well as the sparse-point interpolation routines required by seismic
inversion operators.

Moreover, we demonstrate that Devito-generated kernels are capable of
exploiting modern high performance computing architectures by
achieving a significant percentage of machine peak. Devito's
code-generation engines achieve this by automating well-known
performance optimizations, as well as domain-specific optimizations,
such as flop reduction techniques - all while maintaining full
compatibility with the scientific software stack available through the
open-source Python ecosystem.

\subsubsection{Limitations and Future Work%
  \label{limitations-and-future-work}%
}

The examples used in this paper have been chosen for their relative
simplicity in order to concisely demonstrate the current features of
the Devito API. Different numerical methods may be used to solve the
presented examples with greater accuracy or achieve more realistic
results. Nevertheless, finite difference methods play an important
role and are widely used in academic and industrial research due to
the relative ease of implementation, verification/validation and high
computational efficiency, which is of particular importance for
inversion methods that require fast and robust high-order PDE solvers.

The interfaces provided by Devito are intended to create
high-performance operators with relative ease and thus increase user
productivity. Several future extensions are planned to enhance the
high-level API to further ease the construction of more complex
operators, including explicit abstractions for symbolic boundary
conditions, perfectly matched layer (PML) methods and staggered grids.
Devito's secondary low-level API and use of several intermediate
representations are intended to ease the gradual addition of new
high-level features.

Moreover, the addition of YASK as an alternative backend will not only
provide more advanced performance optimisation, but also an MPI
infrastructure to allow Devito to utilise distribute computing
environments. Further plans also exist for integration with linear
and non-linear solver libraries, such as PETSc, to enable Devito to
handle implicit formulations.

\subsection{Acknowledgements%
  \label{acknowledgements}%
}

This work was financially supported in part by EPSRC grant
EP/L000407/1 and the Imperial College London Intel Parallel Computing
Centre. This research was carried out as part of the SINBAD project
with the support of the member organizations of the SINBAD Consortium.
Part of this work was supported by the U.S. Department of Energy,
Office of Science, Office of Advanced Scientific Computing Research,
Applied Mathematics and Computer Science programs under contract
number DE-AC02-06CH11357.”

\end{document}